\documentclass[%
 aip,pof,
 amsmath,amssymb,
 reprint,%
]{revtex4-1}

\usepackage{graphicx}% Include figure files
\usepackage{dcolumn}% Align table columns on decimal point
\usepackage{bm}% bold math
\usepackage[utf8]{inputenc}
\usepackage[T1]{fontenc}
\usepackage{mathptmx}
\usepackage{etoolbox}

\makeatletter
\def\@email#1#2{%
 \endgroup
 \patchcmd{\titleblock@produce}
  {\frontmatter@RRAPformat}
  {\frontmatter@RRAPformat{\produce@RRAP{*#1\href{mailto:#2}{#2}}}\frontmatter@RRAPformat}
  {}{}
}%
\makeatother
\begin{document}

\preprint{AIP/123-QED}

\title[Rayleigh breakup]{A revision on Rayleigh capillary jet breakup}
% Force line breaks with \\
\author{Alfonso M. Ga{\~n\'a}n-Calvo}
\email{amgc@us.es}
\affiliation{Dept. Ing. Aerospacial y Mec{\'a}nica de Fluidos,
Universidad de Sevilla.\\
Camino de los Descubrimientos s/n 41092, Spain.}
\affiliation{Laboratory of Engineering for Energy and Environmental Sustainability, Universidad de Sevilla, 41092, Spain.}

\date{\today}% It is always \today, today,

\begin{abstract}
The average Rayleigh capillary breakup length of a cylindrical Newtonian viscous liquid jet moving with homogeneous velocity $\hat{U}$ (negligible external forces) must be determined by the selection of normal modes with time-independent amplitude and wavelength (invariant modes, IMs). Both positive and negative group velocity IMs exist in ample ranges of the parameter domain (Weber and Ohnesorge numbers), which explains (i) the average breakup length independence on ambient conditions (long-term resonance), and (ii) its proportionality to the inverse of the spatial growth rate of the dominant positive group velocity IM. Published experimental results since Grace (1965, PhD Thesis) confirm our proposal.
\end{abstract}


\maketitle

Capillary viscous liquid jets are ubiquitous fluid structures in nature and technology, with an immense literature devoted to them (we save space and time to the reader by referring to a couple of reviews\cite{EV08,MG20}). Their intrinsically unstable nature, leading to their eventual breakup into droplets, can be rigorously studied via analytical linear instability analysis whenever their basic unperturbed state with negligible external forces (for example, when gravity is negligible compared to capillary forces) can be consistently reduced to an infinite capillary cylindrical Newtonian liquid column of radius $R$. The normal mode analysis with varicose ($m=0$) perturbations of the form $e^{\text{i}(k z -\Omega t)}$ leads to a dispersion relationship that can be canonically expressed as \cite{W31,C61}:
\begin{align}
   \Delta (\Omega,k) = (k^2+\chi^2)^2 \frac{I_0(k)}{I_1(k)}
    -4k^3\chi\frac{I_0(\chi)}{I_1(\chi)}+\nonumber\\
    k\left(2(k^2-\chi^2)+(k^2-1)/\text{Oh}^2\right) = 0,
    \label{e1}
\end{align}
with $\chi=\left(k^2-i \Omega/\text{Oh}\right)^{1/2}$, where $Oh=\mu/(\rho\sigma R)^{1/2}$ is the Ohnesorge number, and $\rho$, $\mu$ and $\sigma$ are the liquid density, viscosity and surface tension, respectively. Lengths and times are made dimensionless with $R$ and $\left(\rho R^3/\sigma\right)^{1/2}$, respectively.

Assuming that the liquid column is moving at a homogeneous speed $\hat{U}$, Doppler and spatial growth effects that are absent in (\ref{e1}) must be considered. These effects are incorporated exactly making $\Omega=\omega - k U$ (Keller's transformation  \cite{KRT73}) and considering $\omega=\omega_r+\text{i} \omega_i$ and $k=k_r+\text{i} k_i$ complex in (\ref{e1}) (spatiotemporal instability analysis), where $U=W\!e^{1/2}$ and $W\!e=\rho \hat{U}^2 R/\sigma$. Thus, the dispersion relationship (\ref{e1}) defines a five-dimensional manifold in the general six-dimensional real space $\{Oh, U, \omega_r,\omega_i,k_r,k_i\}$ which comprises the whole normal mode spectrum.

The linear wave nature of spatiotemporal normal modes imply the concept of group velocity\cite{B60}, defined from (\ref{e1}) as\cite{W74}:
\begin{equation}
    W = d\omega/d k= -(\partial_k \Delta)/(\partial_\omega \Delta).
    \label{e2}
\end{equation}
In conservative dispersive media, it coincides with the velocity of energy transport, but not in dissipative media (e.g. with viscous damping) where $W$  is generally complex. Its real and imaginary parts $W=U_g+ \text{i} U_d$ (i.e. $U_g=$ Re$(d\omega/dk)$ and $U_d=$ Im$(d\omega/dk)$) have kinematic interpretations\cite{MD93,GS2010} as the envelope propagation velocity of a wave packet with central wavenumber $k$ and the temporal drift experienced by that wavenumber, respectively. Interestingly, it has been shown\cite{GS2010} that in those dissipative media where $W$ can be real (i.e. those that can exhibit waves whose wavenumber has no temporal drift), $W=U_g$ and coincides with the propagation velocity of energy as in nondissipative media. An infinite cylindrical capillary viscous liquid column moving with uniform speed $U$ is an example of such media, and this work aims to exploit its physical implications.

The group velocity concept \cite{B64,HM90a,S03b} and Keller's transformation  allowed Leib and Goldstein \cite{LG86a} to describe the spatiotemporal convective-absolute (C-A) instability limit in terms of a marginal mode $\{\omega^*,k^*\}$ such that its temporal growth rate and group velocity are zero\cite{HM90a}: $\left.\omega_i\right|_{k^*}=0$, $U_g=$ Re$\left. (d\omega/dk)\right|_{k^*}=0$ and $U_d=$ Im$\left.(d\omega/dk)\right|_{k^*}=0$, respectively. These conditions together with Cauchi-Riemann ones imply that $\left. \partial \omega_r/\partial k_r \right|_{k^*}=\left. \partial \omega_r/\partial k_i \right|_{k^*}=0$ with $\omega_i=0$, from which the classical saddle-point criteria immediately follows. This defines the marginal Leib-Goldstein (L-G) curve $U=U^*(Oh)$ in the $\{Oh,U\}$ domain.

Once in the convective instability domain ($W\!e > W\!e^*$ or $U>U^*$), the observed long-term natural breakup regime of a steady cylindrical viscous capillary jet, with a relatively narrow range of breakup lengths, suggests a causal argument: the {\it average} long-term breakup length must be determined by the normal modes with time-independent local amplitude and wavelength (i.e. {\it invariant modes}, IMs), and positive downstream spatial growth rate. Thus, the IMs must satisfy three fundamental conditions for long-term dominance:

(i) $\omega_i=0$ (time-independent local amplitude),

(ii) Im$(d \omega/d k)=0$ (no spatial drift of its wavelength \cite{MD93,GS2010}), and

(iii) $k_i<0$ (i.e. $-k_i$ positive according to the sign criteria here adopted).

\begin{figure*}[htb]
\centering
\includegraphics[width=0.50\textwidth]{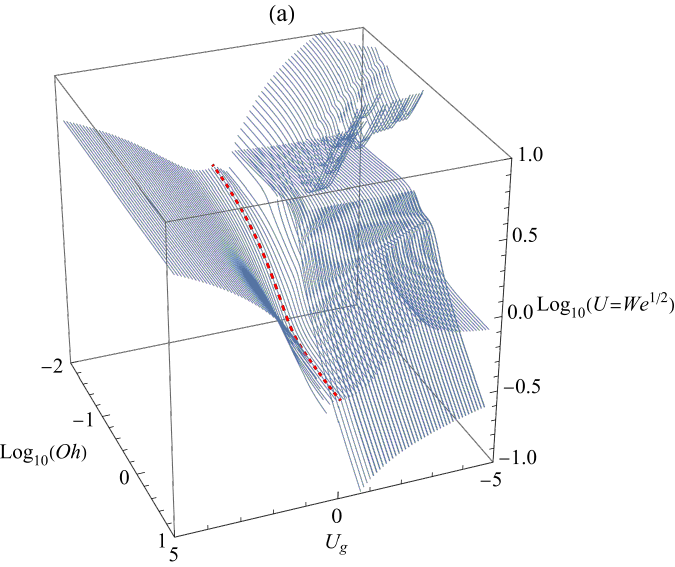}\includegraphics[width=0.50\textwidth]{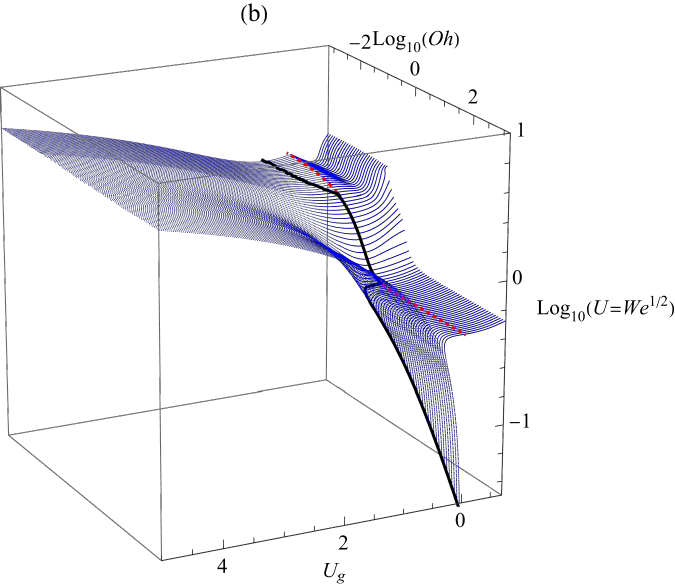}\\
\includegraphics[width=0.7\textwidth]{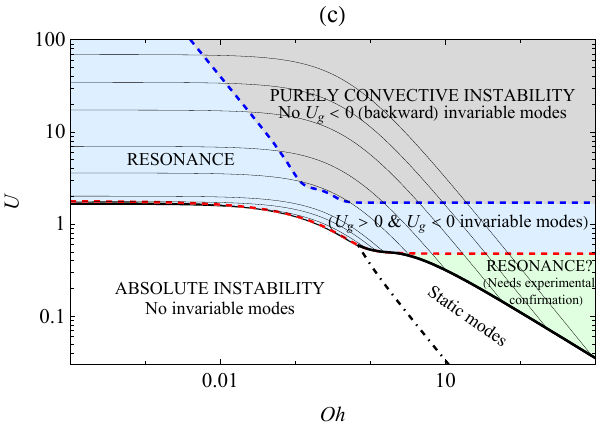}
\vspace{-1mm}
\caption{(a) An illustrative 3D view of the relevant positive and negative group velocity ($U_g=$ Re$(d\omega/dk)$) sheet of invariant modes (IMs), with lines of constant $U_g$ values, here shown as a topologically complex surface in the $\{Oh,U,U_g\}$ space. There is branch-cut topology that connects both sides of the sheet at the central eyelet. The red thick dashed line is the Leib-Goldstein C-A marginal stability limit values $U^*$ ($\omega_i=0$, $d\omega/dk=0$). (b) A detailed 3D view of the relevant side of the sheet corresponding to the dominant invariant mode (DIM, with $U_g>0$), with lines of constant $Oh$ values, showing the existence of multiple modes for a range of $\{Oh,U\}$ values. The black thick line are the minimum $U^{**}$ values below which no forward DIM {\it and} backwards IMs are found. Note the extreme folding of the sheet for $Oh>O(1)$ and $We<O(1)$. (c) Parametrical map of the expected behavior of the cylindrical capillary column form a source, in view of the topology of the IMs' sheet. The red dashed line is the L-G C-A limit. The black thick line are the $U^{**}$ values. The dot-dashed black line delimits the region below $U^{**}$ where only static modes ($\omega_r=k_r=0$) can be found. The blue dashed line gives the limit $U^{***}$ values above which no backward (negative $U_g$) IMs can be found. The thin black lines are the $\{Oh,U\}$ values of the DIMs for constant $-k_i^{-1}=\{3, 4, 5, 10, 20, 50, 100, 200\}$ values, proportional to the jet length $L_j$ (the smaller $L_j$ is the closer to the $U^{**}$ black line).}
\label{fig3}
\end{figure*}

In physical terms, the long-term condition (ii) automatically imply that the group velocity of IMs is their velocity of energy transport. In mathematical terms, the IMs comprise a range (or sub-manifold) of the full modal spectrum of the system. Such a sub-manifold materializes as a four-dimensional sheet in the five-dimensional space of normal modes. However, there are no restrictions on the sign of the group velocity in this sub-manifold. Figure \ref{fig3} represents the locations of the IM sub-manifold in the parameter space.

Figures \ref{fig3}a,b give an illustrative view of the relevant branches of the topologically complex sheet corresponding to the locations of the IMs using the 3D space $\{Oh,U,U_g\}$. Observe that both {\it positive} and {\it negative} group velocity $U_g$ IMs can be found in ample regions of the $\{Oh,U\}$ space. In these cases, the energy can be effectively transported backward and forward along the capillary jet by the IMs, independently of their phase speed $\omega_r/k_r$ and spatial growth rate. A fundamental causal condition in long-term breakup from a fixed source (nozzle) is that the normal mode that is eventually selected, responsible for the average breakup length and size of resulting droplets, must have a {\it positive real group velocity}. We call it the {\it dominant } IM (or DIM).

Provided that backward IMs exist in $\{Oh,U\}$ space, they can propagate upstream the energy from the breakup region, while the forward DIM that determines the most likely rupture length sweeps that energy toward the rupture. In these cases, in the absence of any external energy input (except the steady injection of kinetic and surface energy from the source), the perturbation at the source must not be arbitrary but self-imposed by the breakup dynamics: the only source of perturbations. This comes from the same long-term steadiness assumption applied to the flow of energy along the jet and mode selection. Consequently, when both forward and backward IMs exist, we propose that the breakup length $L_j$ must be

(1) the result of a long-term mechanical {\it resonance}, where the DIM is the main selected breakup mode, and

(2) proportional to the spatial growth rate $-k_i^*$ of the DIM, i.e.
\begin{equation}
    L_j=-C/k_i^*.
\end{equation}

$C$ must be a constant that may depend on the geometry of the jet source, where the flow of backward energy is choked and scattered\cite{LG86b} by the boundary conditions at the nozzle. That was the main claim of a recent work \cite{GCHWKGLCHMB2021} from rather general experimental observations, in line with the suggestions implied by others \cite{BM2021}, in particular from Umemura\cite{Umemura2016,Umemura2020}. In a recent work Liu et al. \cite{LWGHTZD2021} have beautifully proved the validity of previous assumptions by introducing a continuous controlled energy excess by a laser beam aimed at a narrowly selected position of the jet close to the natural breakup point. It should be emphasized that the introduced energy was not oscillatory. The authors show that the system locks-in: the energy introduced is primarily absorbed by the invariant modes that become amplified over the whole the spectrum, and consequently the breakup turns regular. The excited breakup length decreases compared to the natural one \cite{LWGHTZD2021} due to the energy excess put in the DIM.

Thus, the IMs should define the expected linear dynamical behavior form a nozzle. Figure \ref{fig3}c shows that expected behavior in the parametrical space of the problem, $\{Oh,U\}$. In the light gray region delimited by the blue dashed line (say, $U^{***}$ values), no $U_g < 0$ IMs can be found, with a single, well defined DIM. The expected behavior in this region is a convective instability (jetting) in the classical sense, with a high sensitivity of jet length to noise and ambient conditions. In the white region bounded by the thick black line ($U^{**}$ values), no DIMs are found. In other words, the $U^{**}$ values are the minima of the DIM's sheet (figure \ref{fig3}b). Besides, a region of spatially divergent {\it static} modes ($\omega=0$, $k_r=0$, $k_i<0$) can be found in a sub-region bounded by a dot-dashed black line. The latter would not alter the absolute instability (dripping) behavior expected in the whole white region. Between the white and light gray regions, there are (various) $U_g < 0$ IMs, with a well defined DIM. The $U^*$ values corresponding of the location of the IMs with $U_g=0$, first described by \citet{LG86a} (L-G), is given by a red dashed line. Thus, in the light blue region bounded by the blue ($U^{***}$) and red dashed ($U^*$) lines, one can justifiably expect a behavior characterized by a convective instability (jetting) with a self-sustained long-term resonance, and a low sensitivity of the jet length to external noise and ambient conditions. For $Oh$ values between 0.0462 and 2, this line is coincident with the black line corresponding to the $U^{**}$ values. However, for both $Oh<0.0462$ and $Oh>2$, the $U^{**}$ values lay below the $U^*$ ones. The light green region between the thick black and red-dashed lines is characterized by the existence of two close $U_g>0$ IMs, the one with the larger $U_g$ being the DIM (observe the very tight folding of the IM sheet in figure \ref{fig3}b for $Oh > 2$). A single $U_g <0 $ IM is found in this latter region. Consequently, one could expect a convective instability behavior with long-term resonance in this region, and the $U^{**}$ values would establish new stability limits of jetting below the $U^*$ ones. An experimental confirmation is needed for this latter expected behavior, although both $U^*$ and $U^{**}$ are coincident over nearly two orders of magnitude, i.e. for $Oh\in (0.0462,2)$.

{\it Subsequent discussion and consistency of the proposal.-} The group velocity of the DIM (figure $\ref{fig3}$c) exhibits a monotonous dependency with $\{Oh,U\}$ that asymptotically approaches $U$. In contrast, the backward IMs show a greater complexity, with visible edges that degenerate into separate bangs perpendicular to the $\{Oh,U\}$ plane (i.e. their projections on the $\{Oh,U\}$ plane are lines). These features can be observed in the three-dimensional views of figures \ref{fig3}a,b. A fundamental result here is that at least a negative group velocity mode can be found in a continuous subdomain of $\{Oh,U\}$ values with $U^{**} < U < U^{***}$, where $U^{***}$ is the upper limit of $U$ in the $\{Oh,U\}$ space where both forward and backward IMs exist. Interestingly, that line asymptotically approaches $U^{***} \rightarrow 0.4 Oh^{-1}$, or equivalently, $Ca = U \cdot Oh \rightarrow Ca^{***}=U^{***}\cdot Oh=0.4$. However, for $U\gtrsim U^{***}$, no negative group velocity invariant modes can be found: in this regard, the projections of the bangs previously mentioned in figure \ref{fig3}a, which have virtually no area in the $\{Oh,U\}$ plane, do not offer any effective vehicle for backward energy transmission. In this case, although the breakup mode selection criteria are maintained, the sensitivity of the breakup length to external perturbations may significantly increase. Thus, extremely long jets can be expected under carefully maintained ballistic conditions in vacuum, in contrast to the independency of the breakup length on the surrounding gas density in the Rayleigh regime within the subdomain $U^{**} < U < U^{***}$ that was first observed and reported in detail by Grant \cite{G65}, Grant \& Middleman \cite{GM66} and Fenn \& Middleman \cite{FM69}. Noteworthy, in the work of Liu et al.\cite{LWGHTZD2021} the maximum value of $Oh$ is below 0.05, with $U$ below $U^{***}$, and therefore the presence of backward invariant modes was guaranteed in those experiments, leading to resonance.

\begin{figure*}[htb]
\centering
\includegraphics[width=0.45\textwidth]{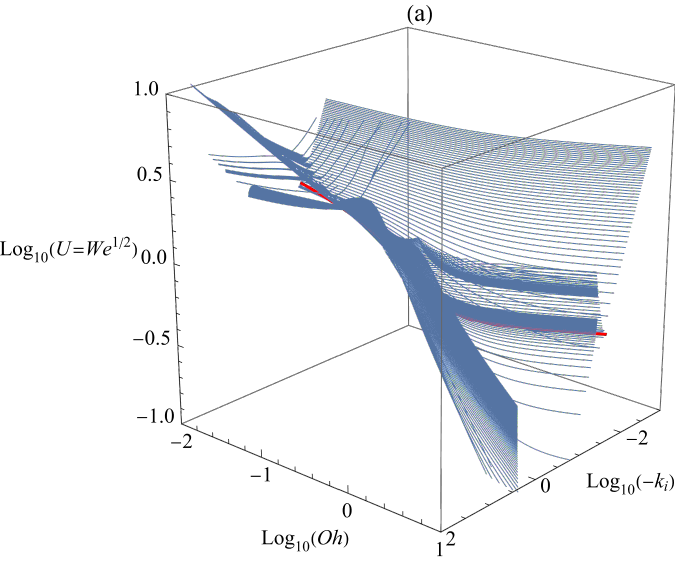}\includegraphics[width=0.45\textwidth]{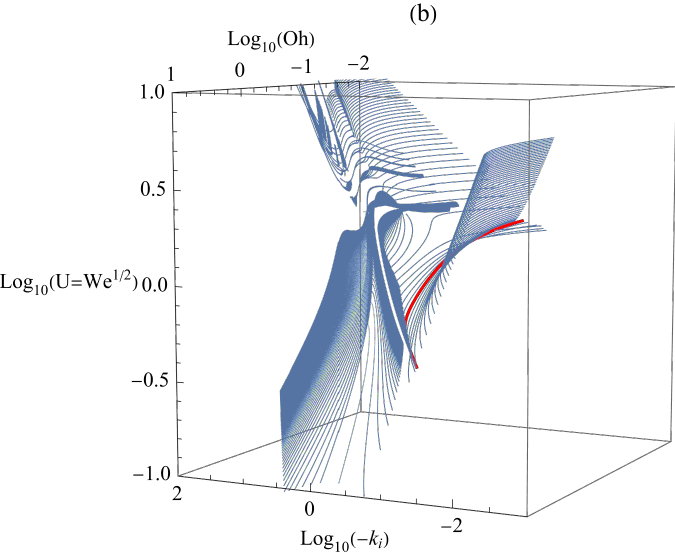}
\vspace{-3mm}
\caption{(a) \& (b) Complete three-dimensional views from different angles of the spatial growth rate $-k_i$ sheet, as a function of $\{Oh,U\}$. The sheet is visualized by the same $U_g$ iso-contours of figure \ref{fig3}. The red line is the C-A marginal stability limit ($d\omega/dk=0$, $\omega_i=0$).}
\label{fig3-2}
\end{figure*}

An additional discussion on the spatial growth $-k_i$ of the IM is necessary. The topological complexity of the IM sheet giving the growth rate values $-k_i$ is apparent in the three dimensional views of figures \ref{fig3-2}a,b. In a first inspection, one observes that the values of $-k_i$ become smaller for the set of DIMs (figure \ref{fig3}c) than the corresponding ones of the backward IMs, for $U^{**} < U < U^{***}$, for $Oh>2$. Thus, the DIMs sweep downstream the energy from farther upstream than the distance of penetration of the corresponding backward IMs. In addition, those DIMs have phase speeds significantly larger than the backward IMs. In consequence, one may expect to find convectively stable viscous jets in the range $U^{**} < U < U^{***}$ with $Oh>2$. In reality, this does not contradicts the analysis carried out by Leib and Goldstein\cite{LG86a} since their work was restricted to Oh values in the range $0.008 < Oh < 0.164$. Further careful experimental work in vacuum is needed to confirm whether the actual stability limit for $Oh>2$ would be $U^{**}$ instead of the marginal C-A limit $U^*$ resulting from the original Leib and Goldstein's proposal $d\omega/dk=0$ (i.e. $U_g=0$ and Im$d\omega/dk=0$) for the whole range of $Oh$ values.

Another fundamental feature of the IMs is that their $-k_i$ reach extreme values in the $\omega_i=0$ spectrum. This is demonstrated in the Appendix A.
First, causal arguments (boundary conditions at the source) lead to consider negative $k_i$ values (i.e. positive growth rate in the downstream direction) and decaying or zero temporal growth $\omega_i\le 0$ modes only. Simple inspection shows that $k_i$ is minimum (maximum $-k_i$ growth rate, see figure \ref{sheets}, Appendix A) for the DIM ($U_d=0$, $U_g>0$). In contrast, the growth rate $-k_i$ is minimum for backward IMs. This means that backward IMs drive energy farther upstream (smaller $-k_i$) that the modes with non-zero Im$(d\omega/d k)$ (i.e. modes with a non-zero temporal drift of their wavenumber). Even more interestingly, one can verify that the whole backward mode spectrum have temporal decaying modes ($\omega_i<0$) with spatial growth $-k_i$ smaller than the locally long-term invariant ones ($\omega_i=0$), independently of their propagation speed $\omega_r/k_r$. Therefore, those latter modes penetrate farther upstream conveying the unsteady energy from the breakup region.

In addition, the only modes with larger spatial growth rate than the DIM are all temporally decaying modes ($\omega_i<0$). Interestingly, one finds for the same initial amplitude that $e^{(-k_i + \omega_i U)z} > e^{-k_i^* z}$, i.e. $-k_i+k_i^*$ is slightly larger than $-\omega_i/U$ for these decaying modes. However, they cannot overcome the DIM in the long run since the backward energy from each breakup event that would feed those modes is injected during a short fraction of the average breakup period. In fact, the existence of these spatially faster growing but temporally decaying modes is the only reason why a rather {\it chaotic} short-term natural breakup is observed, rather than a perfectly regular breakup as one would expect from a short-term perfect resonance with a single overall dominant mode\cite{LWGHTZD2021}. This is a beautiful illustration of the maintenance of a certain level of chaos in many natural mesoscale processes.
This is also the reason why a regular breakup is possible with excitation frequencies $\omega_r$ different from the one of the DIM, $\omega_r^*$, if the excitation energy (or initial amplitude) is large enough for that specific mode \cite{Be88,AM95,LG04,GG08,GG09,GGCC14}. The only possible short-term resonance with an non-oscillating excitation must be in those conditions when the continuous energy input can be mainly absorbed by the DIM, like in the experiments of Liu et al. \cite{LWGHTZD2021}. %The consistency of these observations makes the self-sustained long-term (average) invariant breakup hypothesis for $U^{**}< U < U^{***}$ hardly refutable, even in the total absence of any outer noise or thermal fluctuations \cite{Umemura2016,GCHWKGLCHMB2021}.

It is important to note that the values of the spatial wavenumber $k_r^*$ of the DIM are close (but {\it not equal}\cite{LG86b}) to the ones of the more classical and simpler temporal stability analysis \cite{Rayleigh18784,C61} (with no $U$ involved). This justifies the success of the latter and the applicability of the Rayleigh prediction to the expected droplet size at breakup. In fact, the spatial growth rate $-k_i^*$ of the DIM can be expressed as $-k_i^*=\Omega_i^*/U$ using Keller's transformation with $\omega_i=0$, where $\Omega^*$ would be the observed temporal growth rate of the DIM in a fixed frame where the jet moves with speed $U$. However, $\Omega_i^*$ does not coincide with the maximum temporal growth rate $\Omega_i^{max}$ of the temporal analysis, except in the $U\rightarrow \infty$ limit, as noted by Leib and Goldstein\cite{LG86b}: note that the temporal analysis implicitly assumes $k_i=0$ (i.e. temporal analysis assumes a real $k$ in (\ref{e1}), only).

\begin{figure*}[htb]
\centering
\includegraphics[width=0.65\textwidth]{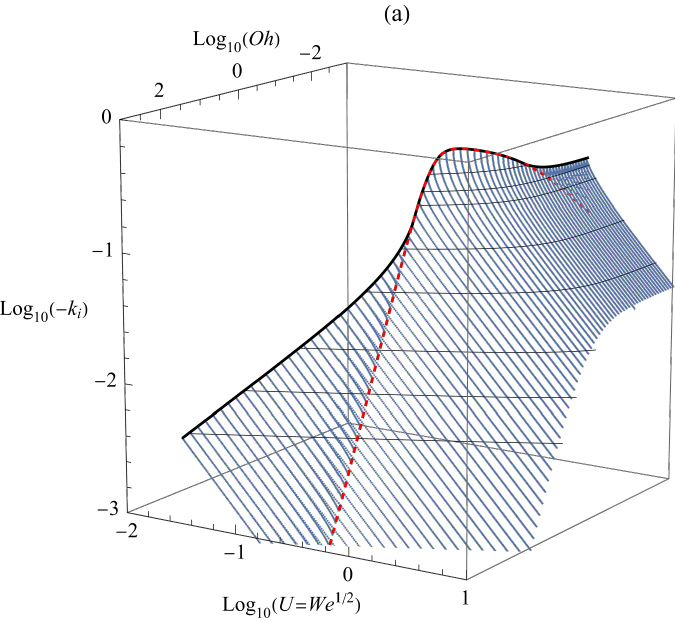}
\vspace{-3mm}
\caption{Three dimensional view showing the values of the spatial growth rate $-k_i^*$ for the positive group velocity $U_g>0$ dominant mode (DIM), using both $Oh-$constant (blue thin lines), and $-k_i^{-1}=\{3, 4, 5, 10, 20, 50, 100, 200\}$ (thin black lines) iso-contours. The black and red dashed thick lines are the $-k_i^*$ values of the DIMs complying $U=U^{**}$, and $\omega_i=0$ \& $d\omega/dk=0$ (Leib \& Goldstein C-A marginal stability limit), respectively.}
\label{fig4}
\end{figure*}

{\it Experimental validation of the proposal.- } To perform an efficient comparison with experiments, note that the two-dimensional dependence of the DIMS's wavenumber $k_i=k_i^*(Oh,U)$, shown in figure \ref{fig4}, exhibits an interesting regularity that was already suggested in the temporal analysis of Weber \cite{W31}. In fact, a single variable $\zeta=2 U\left(1+c_\mu Oh\right)$ was already present in that temporal analysis: the optimum wavenumber (very close to that of the spatiotemporal DIM) and the maximum temporal growth can be expressed under a rather general approximation as $k_r^{opt}=\zeta^{-1/2}$ and $\Omega_i^{max}=c_\omega^{-1} \zeta^{-1}$, with $c_\mu=(9/2)^{1/2}$ and $c_\omega=2^{1/2}$. Translated to our spatiotemporal analysis with $\omega_i=0$, and assuming $\Omega_i^{max}\simeq \Omega_i^*$, one can express $\Omega_i^*=-U k_i^*\simeq 2^{-1/2}\zeta^{-1}$ (the larger $U$, the more exact the latter).

\begin{figure*}[htb]
\centering
\includegraphics[width=0.50\textwidth]{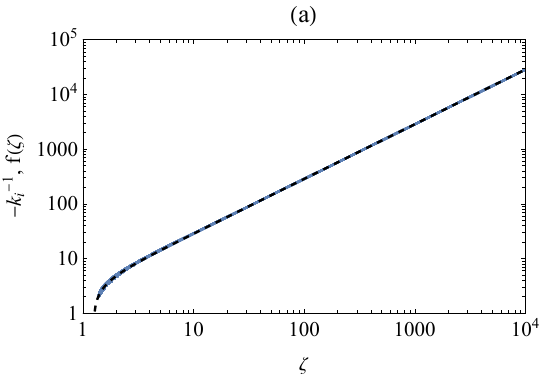}\includegraphics[width=0.50\textwidth]{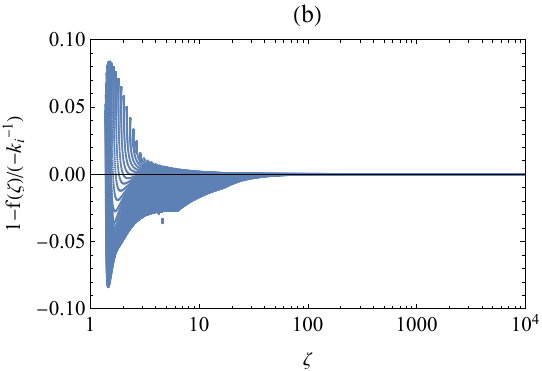}\\
\vspace{-3mm}
\caption{(a) Values of the approximate function $f(\zeta)$, with $\zeta=2 U(1+c_\mu Oh)$ (black dashed line), compared to the $(-k_i^*)^{-1}$ values in the range $10^{-3} < Oh < 10^{3}$ and $W\! e^*< W\! e<40$. (b) Relative errors between $(-k_i^*)^{-1}$ values and $f(\zeta)$.}
\label{fig5}
\end{figure*}

Actually, the reduction of the dependency $-k_i^{-1}=f(Oh,U)$ to a single-variable one as $-k_i^{-1}=f(\zeta)$ with $\zeta=2 U(1+c_\mu Oh)$ was used in many subsequent works (see\cite{GM66,FM69,GCHWKGLCHMB2021}, among many others) to approximate the breakup length $L_j = -C k_i^{-1}$, with $c_\mu$ as a fitting constant. Although we stress again that the DIM does not coincide with the optimum temporal mode with maximum growth rate, the approximations introduced by Weber and the use of a single variable $\zeta$ are sufficiently good to collapse all points shown in figure \ref{fig4} into an approximately single curve. The simultaneous best collapse and best fit (black dashed line) to an expression of the form $f(\zeta)=c_\omega \zeta\left(1-(\zeta_0/\zeta)^{\delta}\right)^{1/\delta}$ with Weber's $c_\mu=(9/2)^{1/2}$ and $c_\omega=2^{1/2}$ gives $\delta\simeq 2.33$ and $\zeta_0\simeq 2.5$ using least squares and maximum regression parameter $R^2$ in the range $W\! e^*< W\! e<40$. The resulting optimum collapse is shown in Figure \ref{fig5}a.
Figure \ref{fig5}b shows the relative errors committed by the obtained fitting function $f(\zeta)$.

\begin{figure*}[htb]
\centering
\includegraphics[width=0.75\textwidth]{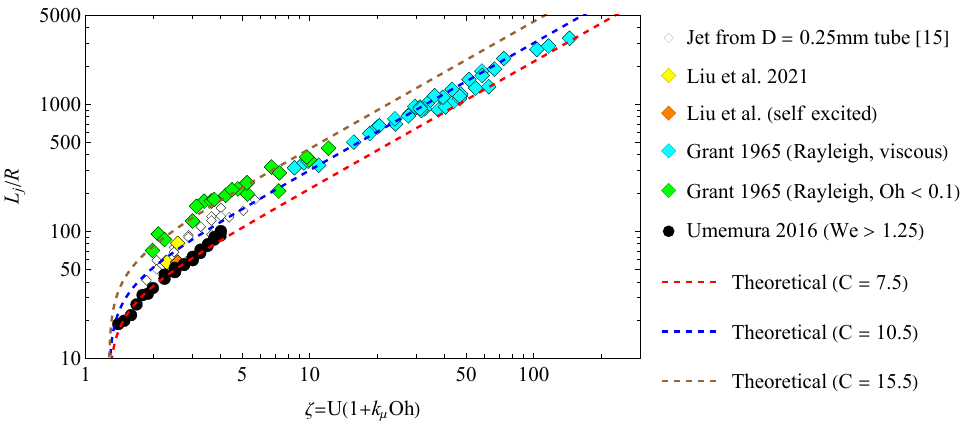}
\vspace{-3mm}
\caption{Experimental data from compared to function $f(\zeta)\times C$, for two values of $C$.}
\label{fig6}
\end{figure*}

In figure \ref{fig6}, we compare our predictions with (i) the classical experiments from \cite{G65} (atmospheric conditions), (ii) the recently published detailed experiments in microgravity of Umemura \cite{Umemura2016}, and (iii) those in \cite{GCHWKGLCHMB2021} for jets from a tube. Two data from Liu et al. \cite{LWGHTZD2021} for natural and self-excited jets from a tube are also shown. The physical properties of the liquids used, nozzle diameters and geometries, etc. can be seen in the respective works and are not given here for economy. The theoretical curves with constants $C=7.5$, 10.5 and 15.5 are plotted for comparison. The classical results of Grant \cite{G65} (summarized in \cite{GM66}) are of particular interest here: first, he reported the insensitivity of the Rayleigh breakup length to the external atmosphere in a parametric range that can be checked to be within $U^{**} < U < U^{***}$, and second, a dramatic increase in sensitivity with the external atmosphere was observed for $Oh \gtrsim O(1)$ and $U>U^{***}$, confirming our proposal.

While an excellent agreement is seen with the data of Umemura \cite{Umemura2016} for $C=7.5$, where the plug velocity is reached much earlier than in the tube experiments, the data from Grant \cite{G65} (jets from capillary tubes) exhibit interesting features: (i) the constant $C$ is larger for tubes than for the orifice experiments, and (ii) $C$ is larger for smaller viscosities. The explanation lies in the jet profile along the axial coordinate, which is clearly convergent in the experiments of Grant during a significant fraction of the total jet length until it reaches a nearly homogeneous (ballistic) diameter and velocity profiles. That homogeneity is reached earlier when the viscosity is larger, which explains the smaller $C$ in those cases. The largest values of $C$ correspond to low viscosity jets for which the initial relaxation length can be above 40\% of the total intact length. If this initial length could be systematically excluded, the predicted length of the cylindrical jet here proposed would approach the experimental measurements of Umemura \cite{Umemura2016}. However, this analysis implies a systematic determination of the zero-th order steady solution as function of $Oh$ and $U$ beyond the scope of present work.

\begin{acknowledgments}

This work was supported by the Ministerio de Econom{\'\i}a y Competitividad (Spain) (project PID2019-108278RB) and the Junta de Andalucía (project P18-FR-3375). Lengthy and enriching discussions with profs. M. A. Herrada, J. M. López-Herrera and J. Eggers are gratefully acknowledged.

\end{acknowledgments}

\section*{Data Availability Statement}

The data that support the findings of this study are available from the corresponding author upon reasonable request.

\appendix

\section{}

\begin{figure}[htb]
\centering
\includegraphics[width=0.40\textwidth]{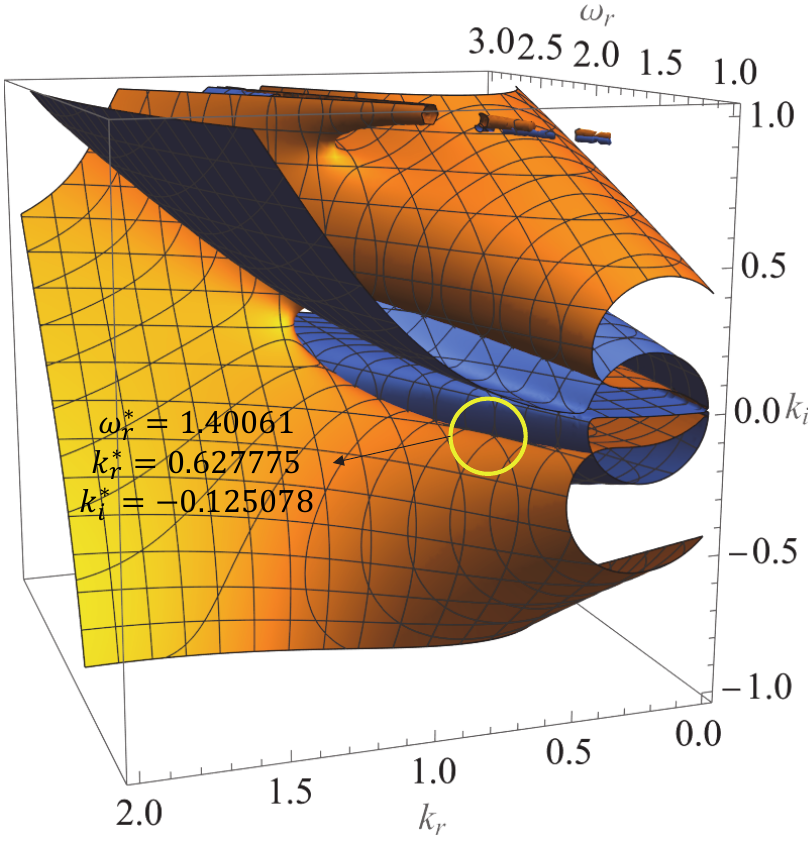}
\vspace{-3mm}
\caption{Sheets of Re$(\Delta)=0$ (blue surface) and Im$(\Delta)=0$ (orange surface), with $\omega_i=0$, for $Oh=0.15$ and $W\!e=U^2=5$. The intersection lines are the modes with zero local variation of their amplitude. Marked at the center of the yellow circle is the mode where $U_d=$ Im$(d\omega/dk)=0$ (DIM), with its corresponding values of $\omega_r^*$, $k_r^*$ and $k_i^*$. The horizontal lines ($k_i=const.$) of the orange surface indicate that $-k_i^*$ is maximum ($k_i^*$ minimum) at that point.}
\label{sheets}
\end{figure}

Current understanding is that the long-term breakup mode selection is determined by the {\it spatial} instability analysis, which restricts the mode spectrum to those modes with $\omega_i=0$ (e.g. \cite{SLYY10}). The dominant mode selection immediately follows from the physical consideration of the maximum (downstream) spatial growth rate $-k_i^{max,\, spatial}$. The rigorous form of the {\it spatiotemporal} mode selection criterion here proposed, i.e. the positive group velocity mode with $\omega_i=0$ and $U_d=Im(d\omega/d k)=0$, leads to the same result exactly, i.e. $k_i^{max,\, spatial}=k_i^*$. In effect, equation (\ref{e1}) implies that
\begin{align}
d\omega/d k= -(\partial_k \Delta)/(\partial_\omega \Delta) = \nonumber \\
\frac{-(\partial_{k_r} \Delta_r \partial_{\omega_r} \Delta_r + \partial_{k_r} \Delta_i \partial_{\omega_r} \Delta_i)}{|\partial_{\omega_r} \Delta|^2} \nonumber \\
+\text{i} \frac{(\partial_{k_r} \Delta_r \partial_{\omega_r} \Delta_i + \partial_{k_r} \Delta_i \partial_{\omega_r} \Delta_r)}{|\partial_{\omega_r} \Delta|^2},
\end{align}
with the standard meaning of subindexes for partial derivatives, and $|\,|$ is the argument. The condition $U_d=$ Im$(d\omega/d k)=0$ leads to
\begin{equation}
\partial_{k_r} \Delta_r/\partial_{\omega_r} \Delta_r = \partial_{k_r} \Delta_i/\partial_{\omega_r} \Delta_i.
\label{c1}
\end{equation}
Since $\omega_i=0$, the real and imaginary sheets of $\Delta=0$ can be visualized in the $\{\omega_r,k_r,k_i\}$ space (see figure \ref{sheets} for $Oh=0.15$ and $W\!e=5$).
In this 3D space, equation (\ref{c1}) implies that the vector product of the normal vectors to both sheets (i.e. $\{\partial_{\omega_r} \Delta_r, \partial_{k_r} \Delta_r,\partial_{k_r} \Delta_r\} \times \{\partial_{\omega_r} \Delta_i, \partial_{k_r} \Delta_i,\partial_{k_r} \Delta_i\}$) has a null component in the $k_i-$direction at the point of the intersection curve $k_i=k_i(\omega_r, k_r)$ where $U_d=$ Im$(d\omega/d k)=0$. Thus, $k_i$ has an extreme value at that point.

%\nocite{*}
%\bibliography{central,central2,biblio}


\end{document}